\documentclass{aa}
\usepackage{amsmath}
\usepackage{txfonts}
\usepackage{braket}
\usepackage{amssymb}
\usepackage{enumerate}
\usepackage{graphicx}
\usepackage{subcaption}
\usepackage{gensymb}
\usepackage{wasysym}

\begin{document}
\title{Detecting chiral asymmetry in the interstellar medium using propylene-oxide}
\titlerunning{Circular dichroism properties of propylene-oxide}
\author{Boy Lankhaar}
\institute{Department of Space, Earth and Environment, Chalmers University of Technology, Onsala Space Observatory, 439 92 Onsala, Sweden \\ 
Leiden Observatory, Leiden University, Post Office Box 9513, 2300 RA Leiden, Netherlands \\
\email{boy.lankhaar@chalmers.se}}

\date{Received 17 June 2022 / Accepted 6 July 2022}

\abstract
   {Life is distinctly homochiral. The origins of this homochirality are under active debate. Recently, propylene-oxide has been detected in the gas-phase interstellar medium (ISM) \citep{mcguire:16}. The enantiomeric composition of ISM propylene-oxide may be probed through circular polarization measurements, but accurate estimates of the circular dichroism properties of the microwave transitions of propylene-oxide are not available.}
   {We aim to develop a model of the circular dichroic activity in torsion-rotation transitions of closed-shell chiral molecules, such as propylene-oxide. With such a model, we estimate the viability, and optimize observation strategies, of enantiomeric excess detection of ISM propylene-oxide.}
   {Circular dichroism in spectral lines manifests through the simultaneous interaction of an electromagnetic radiation field with the molecular electric dipole moment and magnetic dipole moment. We develop techniques to quantify electric dipole and magnetic dipole moments of torsion-rotation transitions by expanding on earlier modeling of \citet{lankhaar:18} on the electric and magnetic dipole properties of single torsion-rotation levels. To model the circular dichroism properties of propylene-oxide, we use these techniques in combination with \textit{ab initio} quantum chemical calculations.} 
   {The expressions for the dichroic activity of the microwave transitions of torsionally active molecules are derived. We find that the torsional motion of molecules exhibiting internal rotation contributes significantly to the total magnetic moment. We present estimates for the dichroic activity of the torsion-rotation transitions of propylene-oxide. We predict that the circular polarization fractions of emission lines of enantiopure propylene-oxide relevant to astronomical detection of propylene-oxide are on the order of $10^{-6}$.}
   {Due to the low predicted circular polarization fractions, we conclude that enantiomeric characterization of propylene-oxide in the gas phase of the ISM is impossible with current astronomical observation techniques. We suggest that only chiral radical species may be viably employed for purposes of enantiomeric excess detection. We estimate that laboratory experiments may be successful in detecting the enantiomeric composition of a mixture of propylene-oxide through microwave dichroism spectroscopy.}
   {}


\maketitle

\section{Introduction}
Homochirality is a key feature of Earth's life, and is hypothesized to be an indicator for extraterrestrial life \citep{avnir:20}. While the origins of the homochirality of life are under active debate \citep{boyd:18}, analysis of carbonaceous chondrites revealed chiral asymmetry in the organic matter \citep{cronin:97}, suggesting that the trait of homochirality that characterizes life on Earth might have cosmochemical origins. The recent detection of propylene-oxide towards Sgt.~B2(N) marked the first chiral molecule to have been discovered in the gas phase of the interstellar medium (ISM) \citep{mcguire:16}. The discovery of the first chiral molecule in the gas phase of the ISM opens up the possibility to probe its possible chiral asymmetry through polarization measurements. 

While the production of ISM propylene-oxide is suggested to yield a racemic mixture \citep{bergantini:18}, chiral asymmetry may be produced through chirally selective photolysis by circularly polarized electrons \citep{ulbricht:62, dreiling:14}, or circularly polarized light \citep{flores:77, modica:14}. A slight chiral asymmetry may be subsequently enhanced through enantioselective surface chemistry \citep{gellman:15}. Models of the chirally selective chemistry in the ISM may be constrained if the enantiomeric composition of propylene-oxide can be probed. 

Enantiomeric excess may be observed through polarization measurements. The transfer of radiation through a non-racemic compound is characterized by the production of circular polarization at its transition frequencies \citep{eyring:68, holzwarth:74}. Chiral compounds that have an enantiomeric excess are dichroic because of the electromagnetic (EM) radiations simultaneous interaction with the molecular electric and magnetic dipole moment \citep{polavarapu:18}. Laboratory measurements of chiral compounds have successfully indicated an enantiomeric excess through the observation of circular dichroism in electronic \citep{eyring:68, caldwell:69} and vibrational \citep{holzwarth:74, stephens:85} transitions. 

Propylene-oxide in the ISM has been detected in its torsion-rotation transitions. While both the electronic \citep{turchini:04, stranges:05, garcia:14, contini:07, rizzo:11} and vibrational \citep{polavarapu:85, kawiecki:88, bloino:12} optical activity of propylene-oxide has been characterized, the circular dichroism of its torsion-rotation transitions has yet to be investigated. In fact, laboratory experiments to the optical activity of any microwave transition are yet to be performed. The circular dichroism properties of rotational transitions have been investigated theoretically \citep{salzman:91, salzman:97}, but these studies did not include the contribution of molecular torsional motion to the optical activity. To determine the viability of the astronomical detection of chiral asymmetry in propylene-oxide, requires accurate theoretical modeling of the dichroic properties of its torsion-rotation transitions. 

In this paper, we investigate the dichroic properties of the torsion-rotation transitions of propylene-oxide and discuss the possibility of detecting a possible enantiomeric excess in the ISM through the partial circular polarization of propylene-oxide's emission. In section 2, we discuss the theory behind the emergence of circular polarization in the emission of a chiral molecule characterized by enantiomeric excess. In section 3, we discuss the torsion-rotation structure of propylene-oxide. Further, we outline the quantum chemical methods we used to compute the relevant electric and magnetic properties of propylene-oxide. We also present the proper theory to compute the dichroic properties of the torsion-rotation transitions of a torsionally active molecule. We present the results of our calculations in section 4. In section 5, we discuss the implications of the dichroic properties of the torsion-rotation transitions for astronomical searches for chiral asymmetry in the ISM gas phase. We conclude in section 6.

\section{Theory}
We set out to evaluate the possibility of detecting enantiomeric excess in propylene-oxide in the ISM through polarization observations. The astronomical emission of propylene-oxide occurs in the microwave region, and is associated with torsion-rotation transitions. To characterize the production of polarization in astronomical propylene-oxide emission, we revise the (polarized) radiative transfer equation through a chiral medium in the following subsection. The production of circular polarization at the transition frequencies of a chiral molecule is the product of the simultaneous interaction of the electromagnetic radiation with the molecular electric dipole and magnetic dipole moments. In section~2.2, we review the theory of the electric dipole and magnetic dipole moment of torsionally active molecules.
\subsection{Radiative transfer and circular dichroism}
The transfer of electromagnetic radiation through a medium with chiral molecules is characterized by the production of circular polarization at the transition frequencies of the chiral molecule, called circular dichroism, and the rotation of linear polarization away from the transition frequencies, called optical rotation. Both of these effects are the product of the simultaneous interaction of the electromagnetic radiation with the molecular electric dipole and magnetic dipole moments, and are related to each other through the Kramers-Kronig relations. Circular dichroism of microwave transitions is the preferred method of astronomical chirality detection, as this effect occurs at the transition frequencies which are specific to the (chiral) molecule, and it does not require a linearly polarized background radiation field. 

In Appendix \ref{sec:rad_trans}, we derive the polarized radiative transfer equation for a chiral molecule. 
There exist other thorough derivations of interaction of radiation with chiral molecules \citep[see, e.g.,][]{polavarapu:18}, but here we present the polarized radiative transfer in a formalism that will be more familiar to astrophysicists. We favor this formalism as it is more suitable to astrophysical conditions, where both the emission and absorption properties of the medium need to be incorporated in the radiative transfer equation.

We consider a transition between lower state, that we denote by its angular momentum, $j_1$, and upper state, $j_2$. The dichroic properties of the transition have their origin in the simultaneous interaction of the electric dipole and magnetic dipole moment with the electromagnetic (EM) field. The relative interaction rate of that interaction, compared to the regular electric dipole interaction rate, is (in SI or CGS units), 
\begin{align}
\label{eq:kuhn}
g_{\mathrm{Kuhn}}^{j_2,j_1} = \frac{2\mathrm{Im}\left(\mu_{j_2,j_1}m_{j_2,j_1}^*\right)}{c|\mu_{j_2,j_1}|^{2}}.
\end{align}
The relative interaction rate is dimensionless and is called Kuhn's dissymmetry factor. Kuhn's dissymmetry factor is dependent on the transition electric and magnetic dipole moments, $\mu_{j_2,j_1}$ and $m_{j_2,j_1}$ and is normalized by the speed of light, $c$. In the method section and Appendix \ref{sec:ang_mom}, we describe how the transition electric and magnetic dipole moments may be computed for a torsion-rotation transition. The equation describing the transfer of (polarized) radiation for a chiral molecule, over the optical depth $\tau_{\nu}$, is derived in Appendix \ref{sec:rad_trans}, and reads
\begin{align}
\label{eq:rad_trans_th}
\frac{d}{d\tau_{\nu}} \begin{pmatrix} I_{\nu} \\ V_{\nu} \end{pmatrix} = -\begin{pmatrix} 1 & g_{\mathrm{Kuhn}}^{j_2,j_1} \\ g_{\mathrm{Kuhn}}^{j_2,j_1} & 1 \end{pmatrix} \begin{pmatrix} I_{\nu} \\ V_{\nu} \end{pmatrix} + S_{\nu} \begin{pmatrix} 1 \\ g_{\mathrm{Kuhn}}^{j_2,j_1} \end{pmatrix},
\end{align}
where $I_{\nu}$ and $V_{\nu}$ are the Stokes parameters at frequency $\nu$, that represent the total (unpolarized) specific intensity, and the circular polarization, respectively. In Eq.~(\ref{eq:rad_trans_th}), $S_{\nu}$ is the source function given in Eq.~(\ref{eq:source}). 

A transition is circularly dichroic if Kuhn's dissymmetry factor, $g_{\mathrm{Kuhn}}^{j_2,j_1}$, is nonzero. From its definition, we readily observe that it is only nonzero for molecules that are chiral: the electric dipole and magnetic dipole moment operators are respectively antisymmetric and symmetric under an inversion operation, so Kuhn's dissymmetry factor is opposite for two enantiomers and cancels out for a non-chiral molecule or a racemic mixture. From the radiative transfer equation, we note that the fractional circular polarization of an optically thin transition approaches its dissymmetry factor in case of a homochiral sample. For an absorption line, the fractional circular polarization approaches Kuhn's dissymmetry factor times the optical depth.
 
\subsection{Electric and magnetic properties of propylene-oxide}
The properties relevant to the circular dichroism of a chiral molecule are the electric dipole and magnetic dipole moments. The electric dipole moment results from the separation of positive and negative charges in the molecule. The magnetic dipole moment of a closed-shell molecule such as propylene-oxide is the result of the differential motion of its charged particles. The electric dipole moment of a molecule may be derived with relative ease from the line strengths of its transitions \citep{townes:55}. The magnetic dipole moment, on the other hand, is dependent on the motion of the nuclei and electrons, and its evaluation involves rigorous modeling of the often cumbersome rotational dynamics of that molecule. In this paper, we are interested in propylene-oxide, which is a torsionally active molecule, exhibiting internal rotational motion. 

The full derivation of the magnetic dipole moment of torsionally active molecules was first presented by \citet{huttner:69} for a molecule with its principal $a$-axis along the torsional-axis. Later, \citet{sutter:70} generalized this to a molecule with an arbitrary torsional-axis. The magnetic dipole moment of a torsionally active molecule may be characterized by the rotational $g$-tensor, $\boldsymbol{g}$, and the torsional $\boldsymbol{b}$-vector, $\boldsymbol{b}$, which are dimensionless tensors that relate the molecular (internal) rotation to its magnetic dipole moment through \citep{sutter:70, lankhaar:18}
\begin{align}
\label{eq:mag_mom}
\hat{\boldsymbol{m}} = -\frac{\mu_N}{\hbar} \boldsymbol{g} \hat{\boldsymbol{J}} - f \frac{\mu_N}{\hbar}\left[\boldsymbol{b} - \boldsymbol{g}\boldsymbol{\lambda} \right] \left( \hat{p}_{\gamma} - \boldsymbol{\rho} \cdot \hat{\boldsymbol{J}} \right),
\end{align}
where $\hat{\boldsymbol{J}}$ and $\hat{p}_{\gamma}$ are the rotational and torsional angular momentum operators, $\mu_N$ and $\hbar$ are the nuclear magneton and the reduced Planck constant, $f$ is a dimensionless factor that depends on the ratio of the moments of inertia of the rotating top and frame, $\boldsymbol{\lambda}$ is the unit vector in the direction of the internal rotation and $\boldsymbol{\rho}$ is a vector dependent on the relative inertia of the internal rotation group and the molecule. 

The magnetic dipole moment of a torsion-rotation level may be evaluated from the rotational $g$-tensor and torsional $\boldsymbol{b}$-vector. While the rotational $g$-tensor may be accurately determined through quantum chemical modeling \citep{flygare:71, sauer:11}, no such methods exist to numerically compute the torsional $\boldsymbol{b}$-vector. There have been some experimental efforts to measure the torsional $\boldsymbol{b}$-vector of molecules exhibiting internal rotation. \citet{engelbrecht:73} determined the torsional $\boldsymbol{b}$-vectors of some simple molecules exhibiting internal rotation. Recently, \citet{takagi:21}, measured the torsional $\boldsymbol{b}$-vector of methanol, a torsionally active molecule whose Zeeman effect may be detected in its strong maser lines \citep{lankhaar:18, vlemmings:08}.

Evaluating the magnetic dipole moment of a torsion-rotation level or transition, requires evaluating Eq.~(\ref{eq:mag_mom}) in a torsion-rotation basis using angular momentum algebraic techniques. The proper theory of evaluating torsion-rotation level specific magnetic dipole moments from the rotational $g$-tensor and torsional $\boldsymbol{b}$-vector has been presented in \citet{lankhaar:18}. In Appendix \ref{sec:ang_mom}, we expand on \citet{lankhaar:18}, and derive the expressions of magnetic dipole and electric dipole moment matrix elements that are general to torsion-rotation transitions. We furthermore dedicate particular attention to the antisymmetric part of the rotational $g$-tensor, because of the sensitivity of Kuhn's dissymmetry parameter to it. To model the antisymmetric part of the rotational $g$-tensor, we follow the early work of \citet{eshbach:52}, who pointed out the necessity to symmetrize the coupling of the magnetic field to the rotational angular momentum. 

\section{Methods}
In the following, we present the methods that we used to compute the dichroic activity properties of propylene-oxide. First, we discuss the modeling of the torsion-rotation structure of propylene-oxide. After that, we discuss the electronic structure methods that we used to obtain the magnetic coupling tensors. In Appendix \ref{sec:ang_mom}, we present proper angular momentum algebra to find their matrix elements in the basis that we used to obtain the torsion-rotation structure. 
\subsection{Torsion-rotation structure of propylene-oxide}
Propylene-oxide is an asymmetric rotor molecule that is torsionally active about its C$-$C${\mathrm{H}_3}$-bond, where the internal rotation is hindered by a three-fold barrier of $947.8$ cm$^{-1}$ \citep{swalen:57}. The internal rotation of the methyl group is associated with a potential with three equivalent minima, resulting in the splitting of each rotational state in two degenerate torsion-rotation states of $E$-symmetry, and one torsion-rotation state of $A$-symmetry. Spectroscopic studies of the microwave spectrum have been performed both in the torsional ground state \citep{swalen:57, mesko:17} and excited torsional state \citep{stahl:21}. The Hamiltonian that we used to model the torsionally active propylene-oxide is described in \citet{swalen:57}. We obtained torsion-rotation eigenstates through a modified internal axis method \citep{woods:66, vacherand:86, hartwig:96}, where the rotational part of the Hamiltonian is set up in the principal axis system, where the inertia tensor is diagonal, while the internal rotation Hamiltonian is set up in the rho axis system \citep{hougen:94} and rotated to be later incorporated into the total Hamiltonian. We used the fitting constants of \citet{mesko:17} to set up the torsion-rotation Hamiltonian of propylene-oxide. The eigenfunctions of the torsion-rotation Hamiltonian were expanded in a basis of $\ket{J(K)M}\ket{v_{\tau} (K) \sigma}$ functions. The functions $\ket{J(K)M}$ are the regular symmetric top wave functions, with angular momentum $J$ and body-fixed and space-fixed projection quantum numbers, $K$ and $M$. These functions are expanded with the torsional functions $\ket{v_{\tau} (K) \sigma}$, that are determined for each $K$, and have the torsional quantum number $v_{\tau}$ and symmetry quantum number $\sigma$. The torsional functions are a linear combination of $e^{i(3q + \sigma)\gamma}$ functions, where $\sigma=0$ corresponds to $A$-symmetry states and $\sigma= \pm 1$ to the (degenerate) $E$-symmetry states, and $q$ is an integer and truncated at $|q_{\mathrm{max}}|$ which we chose $10$. 
\subsection{Electric and magnetic properties of propylene-oxide}
The rotational $g$-tensor was obtained from \textit{ab initio} electronic structure calculations with the program package CFOUR \citep{CFOUR}. Calculations with CFOUR were carried out at the coupled-cluster level of theory including single and double excitation with perturbative treatment the triples contributions [CCSD(T)], in a correlation-consistent polarized triple-zeta (cc-pVTZ) basis set. The rotational $g$-tensor was calculated for 10 torsional angles $\gamma$. We fitted the torsionally dependent rotational $g$-tensor elements to sinoidal functions, 
\begin{align}
\label{eq:gtens_fit}
g_{ij} (\gamma) = \sum_{n=0}^2 \left(a_n \cos 3n\gamma + b_n \sin 3n\gamma \right),
\end{align}
where $\gamma=0$ corresponded to the eclipsed configuration. We obtained excellent fits of the \textit{ab initio} data to the expansion functions. In Table \ref{tab:gtens}, we report the fitting parameters. For most elements the variation with the torsion angle is around $10-20\ \%$ of the torsionally averaged rotational $g$-tensor, while for the $g_{aa}$ and $g_{ac}$-elements, the $\gamma$-variation is larger than the average value. The calculation of the torsional $\boldsymbol{b}$-vector has not been implemented in available quantum chemical program packages. The torsional $\boldsymbol{b}$-vector was estimated following a similar procedure as \citet{lankhaar:18} used for the similar molecule methanol. The torsional $\boldsymbol{b}$-vector was assumed, $\boldsymbol{b}=g_{\gamma} \boldsymbol{\lambda}$, to lie along the internal rotation axis, $\boldsymbol{\lambda}$. The factor $g_{\gamma}$ was put at $0.34$, close to the $g$-factor of a general methyl-group. The $\boldsymbol{\lambda}=(\lambda_a,\lambda_b,\lambda_c)=(0.8843,0.0104,0.4668)$ vector was taken from \citet{mesko:17}. For the electric dipole moment, we used the experimentally determined electric dipole moment vector $ (\mu_a,\mu_b,\mu_c)=(0.95,1.67,0.56)\ \mathrm{D}$ \citep{swalen:57, mesko:17}. The $\boldsymbol{\rho}$-vector and $f$-factor that are relevant to the magnetic interactions and energy-spectrum were taken from \citet{mesko:17} and are $\boldsymbol{\rho}=(\rho_a,\rho_b,\rho_c)=(0.1007,0.0030,0.0173)$ and $f=1.0997$. 
\begin{table*}[h!]
\centering 
\caption{Coefficients $a_n$ and $b_n$ (see Eq.~\ref{eq:gtens_fit}) describing the $\gamma$-dependent rotational $g$-tensor elements calculated \textit{ab initio}. The components of the rotational $g$-tensor are dimensionless, and are defined with respect to the principal axes $a$, $b$ and $c$.} 
\begin{tabular}{l c c c c c}
\hline
 & $a_0$ & $a_1$ & $b_1$ & $a_2$ & $b_2$ \\
\hline \hline
$g_{aa}$ & $-1.250 \times 10^{-3}$ & $-1.326 \times 10^{-3}$  & $9.141 \times 10^{-4}$  & $3.096 \times 10^{-6}$  & $3.096 \times 10^{-6}$ \\
$g_{bb}$ & $-3.067 \times 10^{-3}$ & $1.138 \times 10^{-3}$  & $-8.918 \times 10^{-5}$  & $-7.406 \times 10^{-5}$  & $-7.406 \times 10^{-5}$ \\
$g_{cc}$ & $3.895 \times 10^{-3}$ & $6.323 \times 10^{-4}$  & $-2.804 \times 10^{-4}$  & $-4.408 \times 10^{-5}$  & $-4.408 \times 10^{-5}$ \\
$g_{ab}$ & $1.726 \times 10^{-2}$ & $-6.126 \times 10^{-4}$  & $1.795 \times 10^{-4}$  & $-6.341 \times 10^{-6}$  & $-6.341 \times 10^{-6}$ \\
$g_{ba}$ & $3.306 \times 10^{-2}$ & $-4.181 \times 10^{-4}$  & $4.178 \times 10^{-4}$  & $5.284 \times 10^{-5}$  & $5.284 \times 10^{-5}$ \\
$g_{ac}$ & $2.758 \times 10^{-4}$ & $-6.126 \times 10^{-4}$  & $1.795 \times 10^{-4}$  & $-6.341 \times 10^{-6}$  & $-6.341 \times 10^{-6}$ \\
$g_{ca}$ & $5.171 \times 10^{-2}$ & $-1.836 \times 10^{-3}$  & $5.421 \times 10^{-4}$  & $-1.988 \times 10^{-5}$  & $-1.988 \times 10^{-5}$ \\
$g_{bc}$ & $-1.673 \times 10^{-3}$ & $-3.358 \times 10^{-4}$  & $-4.402 \times 10^{-6}$  & $1.731 \times 10^{-5}$  & $1.731 \times 10^{-5}$ \\
$g_{cb}$ & $-1.874 \times 10^{-3}$ & $-3.762 \times 10^{-4}$  & $-5.026 \times 10^{-6}$  & $1.941 \times 10^{-5}$  & $1.941 \times 10^{-5}$ \\
\hline
\end{tabular}
\label{tab:gtens}
\end{table*}

The values for the electric dipole moment, rotational $g$-tensor and torsional $\boldsymbol{b}$-vector, were used in conjunction with the model on the torsion-rotation structure of propylene-oxide to compute the magnetic and electric dipole moments of the torsion-rotation transitions of propylene oxide. We used Eq.~(\ref{eq:mu_matel}) to compute the reduced electric dipole moment of a torsion-rotation transition, while Eqs.~(\ref{eq:m_matel}) were used to compute the reduced magnetic dipole moments. The dichroic properties of the torsion-rotation transitions were subsequently characterized by computing their dissymmetry factors using Eq.~(\ref{eq:kuhn}). 
\section{Results} 
\citet{mcguire:16} observed three torsion-rotation transitions of propylene-oxide towards Sgt.~B2(N) using the Green Bank Telescope. Through chemical and radiative transfer modeling, \citet{das:19} identified additional potentially observable transitions in ALMA band's $3$ and $4$. Recently, coefficients for the detailed excitation modeling of ISM propylene-oxide has been presented \citep{dzenis:22}. Tables \ref{tab:trans} and \ref{tab:trans_das} give an overview of the line properties and circular dichroism characteristics of the observed microwave transitions and the predicted ALMA band 3 and 4 transitions, respectively. All the investigated transitions are in the torsional ground state. Even though the $A$ and $E$-species of a particular transition are slightly displaced in frequency, we do not report these as individual lines, as the displacement is smaller than typical turbulent broadening in the ISM. We therefore denote transitions by their total rotational angular momentum, $J$, and the projection on the principal axes, $K_a$ and $K_c$: $J_{K_a,K_c}$. Kuhn's dissymmetry factors for both symmetry types, $A$ and $E$, are almost identical in all the investigated transitions. This is a result of the high barrier to the internal rotation. We therefore report a single dissymmetry factor per transition. 
\begin{table*}[h!]
\centering 
\caption{Circular dichroism properties of the propylene-oxide microwave transitions as observed by \citet{mcguire:16}. We report only the line center frequency, but it should be noted that the two symmetry species of each line, $A$ and $E$, are slightly displaced in frequency. The Einstein A-coefficient, $A_{ji}$ is reported for each line, as well as Kuhn's dissymmetry factor, $g_{\mathrm{Kuhn}}$.} 
\begin{tabular}{l c c c}
\hline
transition & $\nu_0$ (GHz) & $A_{ji}$ (s$^{-1}$) & $g_{\mathrm{Kuhn}}$ \\
\hline \hline
$3_{12} \to 3_{03}$ & $14.048$ & $4.157 \times 10^{-8}$ & $2.888 \times 10^{-7}$ \\
$2_{11} \to 2_{02}$ & $12.837$ & $3.327 \times 10^{-8}$ & $2.750 \times 10^{-7}$ \\
$1_{10} \to 1_{01}$ & $12.072$ & $2.854 \times 10^{-8}$ & $2.663 \times 10^{-7}$ \\
\hline
\end{tabular}
\label{tab:trans}
\end{table*}

\begin{table*}[h!]
\centering
\caption{Circular dichroism properties of the propylene-oxide (sub)millimeter transitions in ALMA bands $3$ and $4$ as reported \citet{das:19}. We report only the line center frequency, but it should be noted that the two symmetry species of each line, $A$ and $E$, are slightly displaced in frequency. The Einstein A-coefficient, $A_{ji}$ is reported for each line, as well as Kuhn's dissymmetry factor, $g_{\mathrm{Kuhn}}$.}
\begin{tabular}{l c c c}
\hline
transition & $\nu_0$ (GHz) & $A_{ji}$ (s$^{-1}$) & $g_{\mathrm{Kuhn}}$  \\
\hline \hline
$16_{2\ 15} \to 16_{1\ 16}$ & $85.166$ & $2.464 \times 10^{-6}$ & $1.113 \times 10^{-6}$ \\
$7_{1\ 7} \to 6_{1\ 6}$ & $85.484$ & $2.998 \times 10^{-6}$ & $7.508 \times 10^{-8}$ \\
$17_{4\ 14} \to 17_{3\ 15}$ & $85.856$ & $5.499 \times 10^{-6}$ & $1.137 \times 10^{-6}$ \\
$4_{2\ 2} \to 3_{1\ 3}$ & $88.348$ & $3.861 \times 10^{-6}$ & $5.109 \times 10^{-7}$ \\
$7_{3\ 5} \to 6_{3\ 4}$ & $88.655$ & $2.791 \times 10^{-6}$ & $4.585 \times 10^{-8}$ \\
$7_{4\ 4} \to 6_{4\ 3}$ & $88.599$ & $2.298 \times 10^{-6}$ & $4.417 \times 10^{-8}$ \\
$7_{4\ 3} \to 6_{4\ 2}$ & $88.602$ & $2.299 \times 10^{-6}$ & $4.424 \times 10^{-8}$ \\
$7_{1\ 6} \to 6_{1\ 5}$ & $90.475$ & $3.551 \times 10^{-6}$ & $1.022 \times 10^{-7}$ \\
$17_{2\ 16} \to 17_{1\ 17}$ & $90.723$ & $2.821 \times 10^{-6}$ & $1.178 \times 10^{-6}$ \\
$7_{1\ 7} \to 6_{0\ 6}$ & $91.338$ & $7.928 \times 10^{-6}$ & $1.454 \times 10^{-6}$ \\
$8_{0\ 8} \to 7_{1\ 7}$ & $94.046$ & $8.620 \times 10^{-6}$ & $1.758 \times 10^{-6}$ \\
$3_{3\ 1} \to 2_{2\ 0}$ & $96.421$ & $1.031 \times 10^{-5}$ & $1.224 \times 10^{-7}$ \\
$18_{5\ 13} \to 18_{4\ 14}$ & $97.912$ & $8.298 \times 10^{-6}$ & $1.295 \times 10^{-6}$ \\
$8_{0\ 8} \to 7_{0\ 7}$ & $98.662$ & $4.708 \times 10^{-6}$ & $4.763 \times 10^{-8}$ \\
$8_{2\ 7} \to 7_{2\ 6}$ & $100.647$ & $4.722 \times 10^{-6}$ & $2.712 \times 10^{-8}$ \\
$8_{1\ 8} \to 7_{0\ 7}$ & $102.189$ & $1.161 \times 10^{-5}$ & $1.691 \times 10^{-6}$ \\
$5_{2\ 3} \to 4_{1\ 4}$ & $103.094$ & $4.958 \times 10^{-6}$ & $0.718 \times 10^{-6}$ \\
$14_{5\ 9} \to 14_{4\ 10}$ & $103.205$ & $8.886 \times 10^{-6}$ & $1.352 \times 10^{-6}$ \\
$6_{2\ 5} \to 5_{1\ 4}$ & $105.436$ & $7.160 \times 10^{-6}$ & $1.019 \times 10^{-6}$ \\
$9_{0\ 9} \to 8_{1\ 8}$ & $107.006$ & $1.354 \times 10^{-5}$ & $1.978 \times 10^{-6}$ \\
$4_{3\ 2} \to 3_{2\ 1}$ & $108.983$ & $1.197 \times 10^{-5}$ & $3.460 \times 10^{-7}$ \\
$4_{3\ 1} \to 3_{2\ 2}$ & $109.161$ & $1.201 \times 10^{-5}$ & $3.447 \times 10^{-7}$ \\
$9_{2\ 8} \to 8_{2\ 7}$ & $113.086$ & $6.834 \times 10^{-6}$ & $2.151 \times 10^{-8}$ \\
$9_{1\ 9} \to 8_{0\ 8}$ & $113.153$ & $1.646 \times 10^{-5}$ & $1.927 \times 10^{-6}$ \\
$9_{7\ 2} \to 8_{7\ 1}$ & $113.828$ & $2.902 \times 10^{-6}$ & $4.335 \times 10^{-8}$ \\
$9_{4\ 6} \to 8_{4\ 5}$ & $114.013$ & $5.923 \times 10^{-6}$ & $4.871 \times 10^{-8}$ \\
$9_{4\ 5} \to 8_{4\ 4}$ & $114.032$ & $5.926 \times 10^{-6}$ & $4.922 \times 10^{-8}$ \\
\hline
$17_{6\ 11} \to 17_{5\ 12}$ & $126.748$ & $1.637 \times 10^{-5}$ & $1.631 \times 10^{-6}$ \\
$10_{4\ 6} \to 9_{4\ 5}$ & $126.783$ & $8.571 \times 10^{-6}$ & $5.279 \times 10^{-8}$ \\
$14_{6\ 8} \to 14_{5\ 9}$ & $127.852$ & $1.562 \times 10^{-5}$ & $1.643 \times 10^{-6}$ \\
$15_{6\ 10} \to 15_{5\ 11}$ & $127.659$ & $1.599 \times 10^{-5}$ & $1.641 \times 10^{-6}$ \\
$16_{6\ 11} \to 16_{5\ 12}$ & $127.378$ & $1.626 \times 10^{-5}$ & $1.638 \times 10^{-6}$ \\
$9_{6\ 3} \to 9_{5\ 4}$ & $128.563$ & $1.201 \times 10^{-5}$ & $1.651 \times 10^{-6}$ \\
$10_{6\ 5} \to 10_{5\ 6}$ & $128.488$ & $1.315 \times 10^{-5}$ & $1.650 \times 10^{-6}$ \\
$13_{6\ 7} \to 13_{5\ 8}$ & $128.075$ & $1.521 \times 10^{-5}$ & $1.646 \times 10^{-6}$ \\
$13_{6\ 8} \to 13_{5\ 9}$ & $128.098$ & $1.521 \times 10^{-5}$ & $1.646 \times 10^{-6}$ \\
$12_{6\ 6} \to 12_{5\ 7}$ & $128.249$ & $1.468 \times 10^{-5}$ & $1.647 \times 10^{-6}$ \\
$10_{1\ 9} \to 9_{1\ 8}$ & $128.226$ & $1.039 \times 10^{-5}$ & $0.998 \times 10^{-7}$ \\
$12_{6\ 7} \to 12_{5\ 8}$ & $128.259$ & $1.468 \times 10^{-5}$ & $1.648 \times 10^{-6}$ \\
$10_{6\ 4} \to 10_{5\ 5}$ & $128.486$ & $1.314 \times 10^{-5}$ & $1.650 \times 10^{-6}$ \\
$11_{0\ 11} \to 10_{1\ 10}$ & $132.291$ & $2.801 \times 10^{-5}$ & $2.424 \times 10^{-6}$ \\
$6_{3\ 4} \to 5_{2\ 3}$ & $133.620$ & $1.683 \times 10^{-5}$ & $0.797 \times 10^{-6}$ \\
$11_{9\ 2} \to 10_{9\ 1}$ & $139.105$ & $1.408 \times 10^{-6}$ & $4.403 \times 10^{-8}$ \\
$11_{9\ 3} \to 10_{9\ 2}$ & $139.105$ & $1.408 \times 10^{-6}$ & $4.403 \times 10^{-8}$ \\
$11_{8\ 3} \to 10_{8\ 2}$ & $139.132$ & $6.380 \times 10^{-6}$ & $4.478 \times 10^{-8}$ \\
$11_{8\ 4} \to 10_{8\ 3}$ & $139.132$ & $6.380 \times 10^{-6}$ & $4.478 \times 10^{-8}$ \\
$11_{6\ 6} \to 10_{6\ 5}$ & $139.230$ & $9.535 \times 10^{-6}$ & $4.760 \times 10^{-8}$ \\
$11_{6\ 5} \to 10_{6\ 4}$ & $139.231$ & $9.535 \times 10^{-6}$ & $4.760 \times 10^{-8}$ \\
$11_{4\ 7} \to 10_{4\ 6}$ & $139.567$ & $1.186 \times 10^{-5}$ & $5.732 \times 10^{-8}$ \\
$16_{7\ 10} \to 16_{6\ 11}$ & $151.282$ & $2.545 \times 10^{-5}$ & $1.922 \times 10^{-6}$ \\
$14_{7\ 8} \to 14_{6\ 9}$ & $151.617$ & $2.399 \times 10^{-5}$ & $1.925 \times 10^{-6}$ \\
$13_{7\ 7} \to 13_{6\ 8}$ & $151.740$ & $2.301 \times 10^{-5}$ & $1.927 \times 10^{-6}$ \\
$8_{7\ 1} \to 8_{6\ 2}$ & $152.055$ & $1.231 \times 10^{-5}$ & $1.930 \times 10^{-6}$ \\
$8_{7\ 2} \to 8_{6\ 3}$ & $152.055$ & $1.231 \times 10^{-5}$ & $1.930 \times 10^{-6}$ \\
$12_{5\ 8} \to 11_{5\ 7}$ & $152.058$ & $1.466 \times 10^{-5}$ & $5.346 \times 10^{-8}$ \\
$12_{5\ 7} \to 11_{5\ 6}$ & $152.064$ & $1.467 \times 10^{-5}$ & $5.362 \times 10^{-8}$ \\
$11_{2\ 10} \to 10_{1\ 9}$ & $155.301$ & $2.518 \times 10^{-5}$ & $2.237 \times 10^{-6}$ \\
$13_{1\ 13} \to 12_{0\ 12}$ & $158.512$ & $5.118 \times 10^{-5}$ & $2.860 \times 10^{-6}$ \\ 
\hline
\end{tabular}
\label{tab:trans_das}
\end{table*}

The circular dichroism properties are fully contained in the Kuhn's dissymmetry factor of the transition. An order of magnitude estimate puts the dissymmetry factor on the order of $\sim 10^{-5} $. Most of the torsion-rotation transitions of propylene-oxide have Kuhn's dissymmetry factors that are an order of magnitude smaller than this estimate, due to the large electric dipole moment of propylene-oxide and its relatively small rotational $g$-tensor. 

Dissymmetry factors of the order $10^{-6}$ do not produce any astronomically detectable circular polarization. Only those transitions with electric dipole moments that are more than $3$ orders of magnitude lower compared to strong transitions, transitions that are almost forbidden, show high dissymmetry factors. In Table \ref{tab:pol_frac_forb} we list a range of such lines. Kuhn dissymmetry factors of $\sim 1$ are predicted for these lines, and the emission of an enantio\-pure compound through these transitions is expected to be significantly to completely circularly polarized. However, due to their low transition probabilities, several orders of magnitude lower than regular transition lines, forbidden lines are extremely weak and accordingly undetectable in astrophysical regions.

Because of the minor energy difference between both torsional symmetry states and the relatively high internal rotation barrier of propylene-oxide, we investigated the relative contribution of the internal rotation on the total magnetic dipole moment. We find that the magnitude of the magnetic dipole moment that is due to the torsional motion and the coupling between the torsional and rotational motion, is about half of the total magnetic dipole moment. Even though the internal rotation barrier is high, and the difference between both symmetry states is limited, the contribution of the molecular torsional motion to the magnetic dipole moment is still significant and should therefore be modeled rigorously.

\begin{table}[h!]
\centering
\caption{Circular dichroism properties of some forbidden lines of propylene-oxide (sub)millimeter transitions. We report only the line center frequency, but it should be noted that the two symmetry species of each line, $A$ and $E$, are slightly displaced from each other. The Einstein A-coefficient, $A_{ji}$ is reported for each line, as well as Kuhn's dissymmetry factor, $g_{\mathrm{Kuhn}}$.}
\begin{tabular}{l c c c}
\hline
transition & $\nu_0$ (GHz) & $A_{ji}$ (s$^{-1}$) & $g_{\mathrm{Kuhn}}$ \\
\hline \hline
$2_{2\ 0} \to 2_{0\ 2}$ & $46896.200$ & $1.402 \times 10^{-14}$ & $5.708$ \\
$3_{2\ 1} \to 3_{0\ 3}$ & $47168.500$ & $3.542 \times 10^{-14}$ & $5.697$ \\
$4_{3\ 1} \to 4_{1\ 3}$ & $90187.900$ & $5.972 \times 10^{-13}$ & $3.039$ \\ 
\hline
\end{tabular}
\label{tab:pol_frac_forb}
\end{table}

\section{Discussion}
We investigated the possibility of detecting enantiomeric excess through the measurement of circular dichroism in ISM propylene-oxide. Propylene-oxide in the ISM is observed in its microwave transitions, which are between torsion-rotation states. Up to now, no experiments have been performed to measure and characterize circular dichroism in microwave transitions. Therefore, we characterized the circular dichroism of the microwave transitions of propylene-oxide through theoretical means. This entailed developing the proper theory behind the circular dichroism properties of torsion-rotation transitions, which had not been established before. In the following, we discuss the implications of our results for the possibility of detecting chiral asymmetry in propylene-oxide by means of astronomical circular dichroism measurements and laboratory circular dichroism measurements. Thereafter, we put the developments of the theory behind circular dichroism properties of microwave transitions we present in this paper in the context of previous works. 

We characterized the circular dichroism properties of the torsion-rotation transitions, using quantum-chemical techniques to determine the magnetic properties, in conjunction with accurate modeling of the torsion-rotation structure of propylene-oxide. This method has been shown to be highly accurate when modeling the molecular rotational Zeeman effect \citep{flygare:71, sauer:11}, which is a property that is related to circular dichroism. The torsional contribution to the magnetic dipole moment (and circular dichroism) cannot be computed using available quantum-chemical techniques, but rather was estimated to be close to the magnetic dipole moment of a general torsionally active methyl-group. \citet{engelbrecht:73} measured the torsional magnetic dipole moments due to the internally rotating CH$_3$-groups of nitromethane and methyl-boron-difluoride, and found them to be close (within $5 \%$) to the magnetic dipole moment of methane. Recent work of \citet{takagi:21} showed that for methanol, the torsional magnetic dipole moment associated with its CH$_3$-group is $2/3$ of the magnetic dipole moment of methane. Considering this error-margin, and the relative contribution of the torsion to the total magnetic dipole moment, we conservatively estimate an error of $10\%-15\%$ on our estimates of the circular dichroism properties of the microwave transitions of propylene-oxide. We estimate that ISM propylene-oxide exhibiting an enantiomeric excess, would produce circular polarization fractions in its torsion-rotation transitions on the order of $10^{-6}-10^{-8}$. We find that the circular dichroism of propylene-oxide is too weak for it to be astronomically detectable. With current sensitivity limits on circular polarization fractions of $\sim 10^{-3}$, no currently existing astronomical detection technique exists that can measure such low degrees of circular polarization. 

Even if astronomical polarization instruments were available that could detect circular polarization fractions on the order of $10^{-6}$, the contamination of the circular polarization by the non-paramagnetic Zeeman effect due to the ISM magnetic field would be significant. The circular polarization produced through the Zeeman effect, for a non-paramagnetic molecule such as propylene-oxide, at submillimeter frequencies, with a line width and magnetic field typical of the ISM, is 
\[
\left[p_V\right]_{\mathrm{Zeeman}} \simeq 3.3 \times 10^{-6} \ \bar{g}\left(\frac{\Delta v_{\mathrm{FWHM}}}{1 \ \mathrm{km/s}}\right)^{-1} \left(\frac{\nu_0}{10 \ \mathrm{GHz}}\right)^{-1} \left(\frac{B}{100 \ \mathrm{\mu G}} \right), 
\] 
where $\bar{g}$ is the transition g-factor. It should be mentioned here, that the circular polarization signal from the Zeeman effect is expected to exhibit an $S$-shaped profile, being oppositely polarized in the red and blue part of the line, while the circular polarization due to dichroism is expected to yield an even profile that follows the regular line profile.

Characterizing the enantiomeric composition of the gas-phase ISM may be achieved if a chiral radical species were found. Endowed with a paramagnetic magnetic dipole moment, radicals are significantly more dichroic, with dissymmetry factors of the order $\sim 10^{-2}$. Such dissymmetry factors lead to circularly polarized emission lines of a few percents for mixtures with an enantiomeric excess.

Detecting and characterizing the enantiomeric composition of a compound through circular dichroism measurements is a proven method in laboratory environments. However, such laboratory experiments have only been established for optical and vibrational transitions. Successful experiments to characterize circular dichroism in the (torsion-)rotational transitions---that occur in the microwave region of the EM spectrum---of chiral molecules have not been reported yet. Laboratory experiments should be able to detect the circular polarization fractions of the order $10^{-6}$, that we predict for some torsion-rotation transitions of propylene-oxide. Additionally, any contribution of the Zeeman effect to the circular polarization of spectral lines may be mitigated by experimentally compensating for the Earth's magnetic field. But perhaps, in order to detect and characterize chiral asymmetry through microwave lines, laboratory experiments may defer to the measurement of optical rotation, which is related to the circular dichroism by a Kramers-Kronig transformation. Significant optical rotation on the order of a radian may be achieved at optical depths of $\sim 1/g_{\mathrm{Kuhn}}^{j_2,j_1}$, which should be realizable in laboratory experiments. Alternatively, if an experiment is set up where the microwave transition lines are detected in significant absorption, with optical depths $\gg 1$, this will enhance the polarization fraction to $\sim g_{\mathrm{Kuhn}}^{j_2,j_1} \tau_{\nu}$. 

It has been suggested earlier that non-racemic chiral mixtures may give rise to detectable circular polarization in their microwave transitions \citep{salzman:91, salzman:97}. \citet{salzman:91} predicted through theoretical efforts of similar spirit as described in this paper, for microwave transitions of non-torsionally active molecules, dissymmetry factors that are of the same order of magnitude as the dissymmetry factors that we predict for propylene-oxide. We expanded on the theory that is presented in \citet{salzman:91}, in that (i) we consider the contribution of the torsional magnetic dipole moment, and (ii) we rigorously implemented the proper symmetrization of the antisymmetric contribution to the magnetic dipole moment. Even though the internal rotation of propylene-oxide is hindered by a high-barrier, we find that the effect of the torsional motion on the magnetic properties is significant, contributing close to half of the total magnetic dipole moment. We therefore strongly suggest to rigorously model the torsion-rotation dynamics and its contribution to the magnetic dipole and electric dipole moments, when characterizing the circular dichroism properties of (high-barrier) hindered internal rotor molecules. 

Recently, there have been major advances in the detection of chiral asymmetry in compounds via microwave spectroscopy \citep{patterson:13a,patterson:13b}. Through a three wave-mixing scheme, where the EM waves are resonant with $a$, $b$ and $c$-type transitions \citep{townes:55}, a signal may be generated that is proportional to $\mu_a \mu_b \mu_c$. For a chiral compound, the product $\mu_a \mu_b \mu_c$ is opposite for both enantiomers, which makes this experiment sensitive to an enantiomeric excess \citep{patterson:13a}. Such innovative schemes may be manifested in laboratory experiments, where one can exert control over the phase of an EM wave, but have little utility in astrophysical situations.

\section{Conclusions}
The expressions for the dichroic activity of a torsionally active molecules are derived. Theoretical expressions for the Kuhn's dissymmetry factor of microwave transitions had been derived before \citep{salzman:91}, but we expanded on these earlier modeling efforts by (i) extending the optical activity due to the magnetic dipole moment with the contribution of the torsional motion and (ii) performing the proper symmetrization of the interaction with the magnetic field, as prescribed by \citet{eshbach:52}. We used the newly derived expressions, in combination with \textit{ab initio} modeling of the relevant coupling parameters, to characterize the optical activity of the torsion-rotation transitions of propylene-oxide. The results of our calculations are presented for observed and predicted propylene-oxide transitions in Tables \ref{tab:trans}-\ref{tab:pol_frac_forb}.

With the results of Tables \ref{tab:trans}-\ref{tab:pol_frac_forb}, we predict the fractional circular polarization of a torsion-rotation line due to a enantiomeric imbalance of propylene-oxide. We find that for the range of propylene-oxide lines that have been predicted and observed, the predicted circular polarization fraction is on the order of $10^{-6}-10^{-8}$. Observing such polarization fractions is beyond the current capabilities of modern telescopes. Additionally, circular polarization produced by the Zeeman effect is expected to be of similar magnitude under ISM conditions. Still, both effects may be disentangled due to their different spectral manifestation: while the circular polarization profile due to circular dichroism is expected to follow the line-shape, the Zeeman circular polarization is expected to have an $S$-shaped profile. Forbidden lines may show significant polarization fractions, but the astronomical observation of forbidden transitions in low-abundance species such as propylene-oxide is unfeasible. Detection of enantiomeric excess in the gas-phase ISM with currently available observation techniques may only be achieved using a chiral radical species. 

Laboratory measurements of circular dichroism in vibrational transitions have been able to reach the sensitivity required to detect circular dichroism in microwave transitions \citep{polavarapu:18}. In principle, experiments to directly measure the circular dichroism of microwave transitions should be feasible. Such experiments would be complementary to other detection techniques of chiral asymmetry via microwave spectroscopy \citep{patterson:13a, patterson:13b}. The theory presented in this paper provides a solid theoretical underpinning for circular dichroism measurements in microwave transitions.  

\begin{acknowledgements}
Support for this work was provided by the Swedish Research Council (VR) under grant number 2021-00339. Simulations were performed on resources at the Chalmers Centre for Computational Science and Engineering (C3SE) provided by the Swedish National Infrastructure for Computing (SNIC). Gerrit C. Groenenboom and Ad van der Avoird are acknowledged for helpful comments on a first draft of the manuscript.
\end{acknowledgements}

\bibliography{/Users/boylankhaar/texlibs/lib}

\begin{appendix}
\section{Radiative transfer and circular dichroism}
\label{sec:rad_trans}
In the following, we derive the circularly dichroic radiative transfer equation of chiral molecular spectral lines. In contrast to similar derivations of circular dichroic activity of electronic and vibrational transitions \citep{polavarapu:18}, we work in a formalism that will be more familiar to (radio) astronomers. We favor this formalism as both the emission and absorption properties of the medium are incorporated in the radiative transfer equation. We consider the interaction of an EM wave, composed of an electric field, $\boldsymbol{E}$, and a magnetic field, $\boldsymbol{B}$, with a molecule that possesses an electric and magnetic dipole moment, $\boldsymbol{\mu}$ and $\boldsymbol{m}$. Because the interaction of the EM wave with the electric dipole moment is much stronger than with the magnetic dipole moment, the latter interaction is often ignored. However, without the magnetic field component, light is not chiral, and circular dichroism does not emerge as a feature of the radiation transfer. Therefore, in this work, we will not make that simplification. We note the interaction Hamiltonian of a molecule interacting with an external EM field, in the dipole-approximation \citep{loudon:74}
\begin{align}
\hat{H}_{\mathrm{int}} &= - \boldsymbol{\mu} \cdot \boldsymbol{E} - \boldsymbol{m} \cdot \boldsymbol{B} 
= \left(\mu_{+1} + i\frac{m_{+1}}{c} \right)E_{-1} + \left(\mu_{-1} - i\frac{m_{-1}}{c} \right)E_{+1}, 
\end{align} 
where we have used the relation between the electric and magnetic field components of an EM wave as $\boldsymbol{B} = c^{-1} \hat{\boldsymbol{k}} \times \boldsymbol{E}$, where $\hat{\boldsymbol{k}}$ is the unit vector in the wave propagation direction and $c$ is the speed of light, and where we use CGS units. We give the components of the electric field and molecular properties in a spherical basis $\hat{\boldsymbol{e}}_{\pm 1} = \mp 2^{-\frac{1}{2}}(\hat{\boldsymbol{e}}_{x} \pm i \hat{\boldsymbol{e}}_{y})$ and $\hat{\boldsymbol{e}}_{0} = \hat{\boldsymbol{e}}_{z}$. The unit vector $\hat{\boldsymbol{e}}_{z}$ is chosen along the wave propagation direction, while $\hat{\boldsymbol{e}}_{x}$ and $\hat{\boldsymbol{e}}_{y}$ are perpendicular to each other and to $\hat{\boldsymbol{e}}_{0}$. For the purpose of this elementary derivation, we consider a monochromatic EM wave at (natural) frequency $\omega$ that is traveling along the $\hat{z}$-direction, so that its electric field components are $E_{\pm 1} = \mathrm{Re}[\mathcal{E}_{\pm 1}e^{-i\omega(t-z/c)}]$, where $\mathcal{E}_{\pm 1}$ are the complex amplitudes. 

We consider a transition between two states that are separated in energy by $\hbar \omega_0$, and that have angular momentum, $j_1$, and $j_2$. For our purposes, it is important to dedicate extra attention to the degeneracy of the rotational levels. A level $j$ is split up into $[j]=2j+1$ sublevels denoted by the magnetic quantum number $m$: $\ket{jm}$. So, to resolve the degeneracy of the transition-states, we consider the transitions $\ket{j_1 m_1} \leftrightarrow \ket{j_2 m_2}$, over which we will later average. The interaction between the EM wave and the molecule gives rise to a transition between $j_1$ and $j_2$ when the EM-wave frequency, $\omega$, approaches $\omega_0$. Fermi's Golden rule puts the transition rate at \citep{merzbacher:98}
\begin{align}
\Gamma_{j_1m_1 \to j_2 m_2} &= \frac{\pi}{\hbar^2} \int d\omega \ \left| \Bra{j_2 m_2}\mathcal{E}_{-1}^* (\mu_{+1} + i \frac{m_{+1}}{c} ) \right. \nonumber \\
& \left. + \mathcal{E}_{+1}^* (\mu_{-1} - i \frac{m_{-1}}{c} )\Ket{j_1 m_1}\right|^2 \delta (\omega - \omega_0)). 
\end{align}
In order to obtain matrix elements of the electric dipole and magnetic dipole moments, we use the Wigner-Eckart theorem \citep{biedenharn:81, blum:81}, 
\begin{subequations}
\begin{align}
\braket{j_2 m_2 | \mu_{\pm 1} | j_1 m_1} &= \mu_{j_2,j_1} \sqrt{3} (-1)^{j_2 - m_2} \begin{pmatrix} j_2 & 1 & j_1 \\ -m_2 & \pm 1 & m_1 \end{pmatrix}, \\  
\braket{j_2 m_2 | m_{\pm 1} | j_1 m_1} &= m_{j_2,j_1} \sqrt{3} (-1)^{j_2 - m_2} \begin{pmatrix} j_2 & 1 & j_1 \\ -m_2 & \pm 1 & m_1 \end{pmatrix}, 
\end{align}   
\end{subequations}
where the entity in brackets is the Wigner $3j$-symbol and $\mu_{j_2,j_1}$ and $m_{j_2,j_1}$ are the reduced matrix elements of the electric and magnetic dipole moments. In Appendix \ref{sec:ang_mom}, we derive detailed expressions to quantify the reduced matrix elements for a particular torsion-rotation transition. The rate of absorption from the level $\ket{j_1 m_1}$ to all possible levels in $j_2$, is,
\begin{subequations}
\label{eq:abs_stim}
\begin{align}
\Gamma_{j_1 m_1 \to j_2} &= \sum_{m_2} \Gamma_{\ket{j_1m_1} \to \ket{j_2 m_2}} \nonumber \\
                     &= \frac{\pi}{3 [j_1] \hbar^2} \left\{ \left[|\mu_{j_2,j_1}|^2 + \left|\frac{m_{j_2,j_1}}{c}\right|^2 \right] \left(|\mathcal{E}_{+1} (\nu)|^2 +|\mathcal{E}_{-1} (\nu)|^2 \right) \right. \nonumber \\
& \left. + 2 \mathrm{Im}\left(\frac{\mu_{j_2,j_1} m_{j_2,j_1}^*}{c}\right)\left[|\mathcal{E}_{+1} (\nu)|^2 - |\mathcal{E}_{-1} (\nu)|^2 \right]\right\} \nonumber \\
&= B_{j_1 \to j_2} \left( \left[1 + R_{\mathrm{m.d.}}^{j_2,j_1} \right] I_{\nu} + g_{\mathrm{Kuhn}}^{j_2,j_1} V_{\nu} \right), 
\end{align}
where $R_{\mathrm{m.d.}}^{j_2,j_1} = c^{-2}|m_{j_2,j_1}|^2/|\mu_{j_2,j_1}|^2$ is the ratio between the magnetic and electric dipole moment transition strength, and $g_{\mathrm{Kuhn}}^{j_2,j_1} = 2c^{-1}|\mu_{j_2,j_1}|^{-2}\mathrm{Im}\left(\mu_{j_2,j_1}m_{j_2,j_1}^*\right)$ is Kuhn's dissymmetry factor. The Stokes parameters are related to the electric field amplitudes as $I_{\nu} = c/8\pi (|\mathcal{E}_{+1} (\nu)|^2 +|\mathcal{E}_{-1} (\nu)|^2)$ and $V_{\nu} = c/8\pi (|\mathcal{E}_{+1} (\nu)|^2 -|\mathcal{E}_{-1}(\nu)|^2)$ and are given for frequency $\nu = \omega_0 / 2\pi$. The Einstein B-coefficient is defined as $B_{j_1 \to j_2} = \frac{8 \pi^2}{3 [j_1] \hbar^2 c} |\mu_{j_2,j_1}|^2$. Similar to the rate of absorption, the rate of stimulated emission events can be derived to be
\begin{align}
\Gamma_{j_2 m_2 \to j_1} = B_{j_2 \to j_1} \left( \left[1 + R_{\mathrm{m.d.}}^{j_1,j_2} \right] I_{\nu} + g_{\mathrm{Kuhn}}^{j_1,j_2} V_{\nu} \right),
\end{align}
\end{subequations}
where it may be noted that $g_{\mathrm{Kuhn}}^{j_2,j_1} = g_{\mathrm{Kuhn}}^{j_1,j_2}$. The rate of spontaneous emission bears a close relation to the rate of stimulated emission. 

The interactions of the radiation field with the chiral molecule, whose rates are given in Eqs.~(\ref{eq:abs_stim}), will manifest in the transfer of polarized radiation. It may be noted from Eqs.~(\ref{eq:abs_stim}), that the absorption of Stokes $V$ radiation occurs at a rate of $g_{\mathrm{Kuhn}}^{j_2,j_1}$ times the absorption of Stokes $I$ radiation. Similarly, the production of Stokes $V$ radiation occurs at a rate of $g_{\mathrm{Kuhn}}^{j_2,j_1}$ times the production of Stokes $I$ radiation through both stimulated and spontaneous emission processes. More specifically, the change in flux density, per optical depth, may be noted 
\begin{align}
\label{eq:rad_trans}
\frac{d}{d\tau_{\nu}} \begin{pmatrix} I_{\nu} \\ V_{\nu} \end{pmatrix} = -\begin{pmatrix} 1 & g_{\mathrm{Kuhn}}^{j_2,j_1} \\ g_{\mathrm{Kuhn}}^{j_2,j_1} & 1 \end{pmatrix} \begin{pmatrix} I_{\nu} \\ V_{\nu} \end{pmatrix} + S_{\nu} \begin{pmatrix} 1 \\ g_{\mathrm{Kuhn}}^{j_2,j_1} \end{pmatrix},
\end{align}
where $d\tau_{\nu} = \kappa_{\nu} ds$ is the optical depth over an infinitesimal distance $ds$, and
\begin{align}
\kappa_{\nu} = \frac{h \nu}{4 \pi} B_{j_1 \to j_2} \left( n_{j_1} - \frac{[j_1]}{[j_2]}n_{j_2} \right) \phi_{\nu} \nonumber
\end{align}
is the absorption coefficient \citep{rybicki:08}, dependent on the number densities, $n_j$ of levels $j_1$ and $j_2$ and the line-profile $\phi_{\nu}$. The quantity, 
$S_{\nu}$, is the source function, which is defined by the ratio of the emission and absorption coefficient \citep{rybicki:08},
\begin{align}
\label{eq:source}
S_{\nu} = \frac{2 h\nu^3}{c^2} \left[ \frac{[j_2]n_1}{[j_1]n_2} - 1 \right]^{-1},
\end{align} 
which approaches Planck's function for a thermalized transition.

\section{The electric and magnetic dipole moments of torsion-rotation transitions}
\label{sec:ang_mom}
The electric dipole moment results from the separation of positive and negative charges in the molecule and may be computed as $\boldsymbol{\mu}=\sum_j q_j \boldsymbol{r}_j$, where $q_j$ is the charge of particle $j$ and $\boldsymbol{r}_j$ is its position. The magnetic dipole moment of a closed-shell molecule such as propylene-oxide is the result of the differential motion of its charged particles. The classical definition of the magnetic dipole moment of a set of moving charged particles is \citep{jackson:98}
\begin{align}
\boldsymbol{m} = \frac{1}{2} \sum_j q_j (\boldsymbol{r}_j \times \boldsymbol{v}_j),
\end{align}
where $\boldsymbol{v}_j$ is the velocity of particle $j$. For a molecule such as propylene-oxide, the motion of the nuclei is determined by its rotational motion, as well as the internal rotational motion. \citet{lankhaar:18} derived the magnetic dipole moment of a torsionally active molecule. We re-state here the expressions for the magnetic dipole moment \citep[see, equations 1-2 of][]{lankhaar:18}, but separate these in the following terms
\begin{subequations}
\label{eq:magmom}
\begin{align}
\boldsymbol{m} &= \boldsymbol{m}^{\mathrm{R}} + \boldsymbol{m}^{\mathrm{RT}} + \boldsymbol{m}^{\mathrm{T}}, \\
m_q^{\mathrm{R}} &= \frac{\mu_N}{2\hbar} \sum_{q'} \left( g_{qq'} \hat{J}_{q'} + (-1)^{q'-q} \hat{J}_{q'} g_{q'q} \right), \\
m_q^{\mathrm{T}} &= \frac{f\mu_N}{2\hbar} \left(b'_{q} \hat{p}_{\gamma} + \hat{p}_{\gamma} b'_{q}\right), \\
m_q^{\mathrm{RT}} &= -\frac{f\mu_N}{2\hbar} \sum_{q'} \left( b'_{q} \rho_{q'} \hat{J}_{q'} + (-1)^{q'-q} \hat{J}_{q'}\rho_{q'} b'_{q}  \right),
\end{align}
\end{subequations}
which are analogous to a purely rotational magnetic dipole moment, a purely torsional magnetic dipole moment and the remaining rotation-torsion contribution. The separation in these terms will allow for a clearer treatment of the angular momentum algebra later on. We note in Eqs.~(\ref{eq:magmom}) the spherical element, $q$, of the angular momentum operator $\hat{J}_q$, and the torsional momentum operator $\hat{p}_{\gamma}$ (for a definition, see \citet{lankhaar:18}). The tensors $\boldsymbol{g}$ and $\boldsymbol{b}$ are the (rank-2) rotational $g$-tensor and the (rank-1) torsional $\boldsymbol{b}$-vector \citep{sutter:76, lankhaar:18}. The $\boldsymbol{\rho}$-vector is defined in \citet{lankhaar:18}. The quantity $\mu_N$ is the nuclear magneton, $\hbar$ is the reduced Planck constant and $f$ is a dimensionless factor that depends on the ratio of the moments of inertia of the rotating top and frame \citep[for a definition, see][]{lankhaar:18, hougen:94}. We have symmetrized the magnetic dipole moment interactions of Eqs.~(\ref{eq:magmom}) as prescribed in \citet{eshbach:52}.

We derive the matrix elements of the electric dipole and magnetic dipole moment operators that are relevant to transitions between two states, which are denoted as $J_{K_a K_c}^{A/E} \to J_{K_a' K_c'}^{\prime A/E}$, where $J$ is the total angular momentum, $K_a$ and $K_c$ are the projections of the angular momentum on the principal $a$ and $c$ axes, and $A/E$ denotes the symmetry type. As discussed in section 3.1, the eigenfunctions of the torsion-rotation levels are expanded in a basis of $\ket{j(k)m}\ket{v_{\tau} (k) \sigma}$ functions. In the following, we discuss the matrix elements of the electric and magnetic dipole moment operators in these basis functions. 
\paragraph{Electric dipole moment.}
The general matrix element of the transition electric dipole moment may be factorized  
\begin{align}
\label{eq:mu_matel}
\braket{j_1(k_1)m_1, v_{\tau} (k_1) \sigma|\mu_q|v_{\tau}' (k_2) \sigma,j_2(k_2)m_2} = \nonumber \\
 \braket{v_{\tau} (k_1) \sigma|v_{\tau}' (k_2) \sigma} \mu_{j_1k_1,j_2k_2} \braket{j_1m_1|\hat{T}_{1q}(j_1,j_2)|j_2m_2}, 
\end{align}
because the torsional part does not interact with the electric dipole operator, as we neglected any dependence of it on the torsional angle. We used the Wigner-Eckart theorem for the rotational part, which we formulate in terms of the rank-1 irreducible spherical tensor operator $\hat{T}_{1q} (j_1;j_2)$, which is defined following \citet{blum:81}, by 
\begin{align}
\label{eq:tens_op}
\hat{T}_{LP} (j_1;j_2) = \sum_{m_1 m_2} \ket{j_1 m_1} \bra{j_2 m_2} [L]^{1/2} (-1)^{j_1 - m_1} \begin{pmatrix} j_1 & L & j_2 \\ -m_1 & P & m_2 \end{pmatrix}, 
\end{align}
where $L$ is the rank and $P$ its projection. We note the rotation of the electric dipole moment (rank-1 tensor) between the space-fixed (SF) laboratory frame and body-fixed (BF) principal axis frame as $\mu_q^{\mathrm{SF}}=\sum_k D_{kq}^{(1)*}(\chi \theta \phi) \mu_q^{\mathrm{BF}}$, in terms of Wigner-$D$ matrix elements of the Euler angles $\chi,\  \theta$ and $\phi$. We may express the Wigner-$D$ matrix elements in terms of the irreducible tensor operators of Eq.~(\ref{eq:tens_op}) (see also equation (B5) of \citet{lankhaar:16}). The irreducible matrix element is then readily derived as 
\begin{align}
\label{eq:mu_ir}
\mu_{j_1k_1,j_2k_2} = \sqrt{\frac{[j_1][j_2]}{3}} (-1)^{j_1 - k_1} \sum_k \begin{pmatrix} j_1 & 1 & j_2 \\ -k_1 & k & k_2 \end{pmatrix} \mu_k, 
\end{align}
where the sum over $k$ runs over all principal axis components (in a spherical basis) of the electric dipole tensor.
\paragraph{Magnetic moment.} 
In Eq.~(\ref{eq:magmom}), we formulated the magnetic dipole moment in terms of a purely rotational magnetic dipole moment, torsional magnetic dipole moment and the residual torsion-rotation magnetic dipole moment. We begin by deriving expressions for the matrix element of the pure rotational magnetic dipole moment. We accounted for the dependance of the rotational $g$-tensor on the torsional angle, by expanding it as $g_{qq'} = \sum_{n=-2}^2 e^{in\gamma} g_{qq'}^{(n)}$. We may then factorize, 
\begin{align}
\label{eq:rot_mat}
&\braket{j_1(k_1)m_1, v_{\tau} (k_1) \sigma|m_q^{\mathrm{R}}|v_{\tau}' (k_2) \sigma,j_2(k_2)m_2} =\frac{\mu_N}{2\hbar} \sum_n \braket{v_{\tau} (k_1) \sigma|e^{i n \gamma}|v_{\tau}' (k_2)} \nonumber \\ 
&\times \left(\sum_{q'}\Braket{j_1(k_1)m_1|g_{qq'}^{(n)}\hat{J}_q'+(-1)^{q'-q} \hat{J}_{q'}g_{qq'}^{(n)}|j_2(k_2)m_2} \right).
\end{align}
We note that a rotation of the rotational $g$-tensor $g_{qq'} = \sum_{kk'} D_{qk}^{(1)*} (\chi \theta \phi) g_{kk'}^{\mathrm{BF}}  D_{q'k'}^{(1)} (\chi \theta \phi)$, may be decomposed into a rotation of its 3 irreducible elements ($L=0,1,2$). Coupling these to the rotation operator yields 
\begin{subequations}
\label{eq:rot_mat_both}
\begin{align}
&\braket{j_1(k_1)m_1|\sum_{q'} g_{qq'}^{(n)} \hat{J}_{q'}|j_2 (k_2)m_2} \nonumber \\ &= \sum_L g^{(L),(n)}_{j_1k_1,j_2k_2} \Braket{j_1 m_1|\left[ \hat{\boldsymbol{T}}_{L} \otimes \hat{\boldsymbol{J}} \right]^1_q |j_2 m_2}, \\
&\braket{j_1(k_1)m_1|\sum_{q'} (-1)^{q'-q} \hat{J}_{q'}g_{qq'}^{(n)} |j_2 (k_2)m_2} \nonumber \\ &= \sum_L (-1)^L g^{(L),(n)}_{j_1k_1,j_2k_2} \Braket{j_1 m_1|\left[ \hat{\boldsymbol{J}} \otimes \hat{\boldsymbol{T}}_{L} \right]^1_q |j_2 m_2}, 
\end{align} 
\end{subequations}
where we have used a short-hand notation for the product 
\[
\left[ \hat{\boldsymbol{A}}_{L_1} \otimes \hat{\boldsymbol{B}}_{L_2} \right]^{L}_{M} = \sum_{m_1 m_2} C_{m_1 \ m_2 \ M}^{L_1 \ L_2 \ L} \hat{A}_{L_1 m_1} \hat{B}_{L_2 m_2}, 
\]
of the rotational tensors irreducible elements and where $C_{m_1 \ m_2 \ M}^{L_1 \ L_2 \ L}$ is a Clebsch-Gordan coefficient. The elements of the reduced rotational $g$-tensor read
\begin{align}
\label{eq:g_ir}
g^{(L),(n)}_{j_1k_1,j_2k_2} = (-1)^L \sqrt{\frac{[j_1] [j_2]}{[L]}} (-1)^{j_1 - k_1} \sum_Q \begin{pmatrix} j_1 & L & j_2 \\ -k_1 & Q & k_2 \end{pmatrix} g_{LQ}^{(n)},
\end{align}
where $g_{LQ}^{(n)} = \sum_{kk'} (-1)^{1-k'} C_{-k\ k'\ Q}^{1\ 1\ L} g_{kk'}^{(n)}$ are the irreducible tensor elements of the rotational $g$-tensor in the body-fixed frame. As may be expected, the sum of both parts of $\boldsymbol{m}^{\mathrm{R}}$ cancel each other out at uneven (antisymmetric) $L$. However, it should be noted that the angular momentum operators, $\hat{\boldsymbol{J}}$ and $\hat{\boldsymbol{T}}_L$ do not commute for $L=1$. We note the commutation relation between a tensor operator and the spherical components of the angular momentum operator \citep[see Eq.~(3.210) of][]{biedenharn:81}, $[\hat{J}_{m},\hat{T}_{JM}] = \sqrt{J(J+1)}C_{M\ m\ M+m}^{J\ 1 \ J} \hat{T}_{J,M+m}$, from which we derive
\[
\left[ \hat{\boldsymbol{T}}_L \otimes \hat{\boldsymbol{J}} \right]^1_q - \left[\hat{\boldsymbol{J}} \otimes \hat{\boldsymbol{T}}_L \right]^1_q = -\sqrt{2} \hat{T}_{1q} \delta_{L,1}.
\]
Using angular momentum algebra we may couple the product
\begin{align}
\label{eq:prod_mat}
\braket{j_1 m_1|\left[ \hat{\boldsymbol{T}}_{L} \otimes \hat{\boldsymbol{J}} \right]^1_q |j_2 m_2} &= (-1)^{1+j_1+j_2}\sqrt{[j_2][L]j_2(j_2+1)} \nonumber \\
&\times \begin{Bmatrix}j_1 & j_2 & L \\ 1 & 1 & j_2 \end{Bmatrix} \braket{j_1 m_1 | T_{1q} | j_2 m_2}, 
\end{align}
where the entity between curly brackets is a Wigner $6j$-symbol. Putting Eq.~(\ref{eq:rot_mat})-(\ref{eq:prod_mat}) together, we derive the matrix element of the pure rotational magnetic dipole moment,
\begin{subequations}
\label{eq:m_matel}
\begin{align}
&\braket{j_1(k_1)m_1|m_q^{\mathrm{R}}|j_2 (k_2) m_2} = \frac{\mu_N}{\hbar} \sum_n e^{i n \gamma} \left( \frac{g_{j_1k_1,j_2k_2}^{(1),(n)}}{\sqrt{2}} + (-1)^{j_1 + j_2}\sum_{L=0,2} \right. \nonumber \\
& \left. \times g^{(L),(n)}_{j_1k_1,j_2k_2} \sqrt{[j_2][L]j_2(j_2+1)} \begin{Bmatrix}j_1 & j_2 & L \\ 1 & 1 & j_2 \end{Bmatrix} \right) \braket{j_1 m_1 |T_{1q} | j_2 m_2}. 
\end{align}
The rotation-torsion residual contribution to the magnetic dipole moment has a similar form to the rotational magnetic dipole moment. Replacing the $g$-tensor elements $g_{qq'} \to b_{q}\rho_{q'}$, we may arrive after similar angular momentum algebraic manipulations at the expression, 
\begin{align}
&\braket{j_1(k_1)m_1|m_q^{\mathrm{RT}}|j_2 (k_2) m_2} = \frac{f\mu_N}{\hbar} \sum_n e^{i n \gamma} \left( \frac{\zeta_{j_1k_1,j_2k_2}^{(1),(n)}}{\sqrt{2}} + (-1)^{j_1 + j_2}\sum_{L=0,2} \right. \nonumber \\
& \left. \times \zeta^{(L),(n)}_{j_1k_1,j_2k_2}  \sqrt{[j_2][L]j_2(j_2+1)} \begin{Bmatrix}j_1 & j_2 & L \\ 1 & 1 & j_2 \end{Bmatrix} \right) \braket{j_1 m_1 |T_{1q} | j_2 m_2},
\end{align} 
where the elements $\zeta^{(L),(n)}_{j_1k_1,j_2k_2}$ have the same definition as Eq.~(\ref{eq:g_ir}), only replacing $g_{LQ}^{(n)}$ with $\zeta_{LQ}^{(n)}=\sum_{kk'} C_{k\ k'\ Q}^{1\ 1\ L} \rho_k b_{k'}^{(n)}$.   

Finally, we work out the pure torsional contribution to the magnetic dipole moment. Here, the matrix element may be factorized as
\begin{align}
&\braket{j_1(k_1)m_1,v_{\tau} (k) \sigma |m_q^{\mathrm{T}}|j_2 (k_2) m_2 v_{\tau} (k) \sigma} = \frac{f\mu_N}{2\hbar} \sum_n \nonumber \\ 
&\times b_{j_1k_1,j_2k_2}^{(n)} \braket{v_{\tau} (k) \sigma | \hat{p}_{\gamma}e^{i n \gamma} + e^{i n \gamma} \hat{p}_{\gamma}|v_{\tau}' (k_2) \sigma} \braket{j_1 m_1 |T_{1q} | j_2 m_2},
\end{align}
\end{subequations}
where the Wigner-Eckart theorem for the rotational part of the operator is of a similar form as the electric dipole operator, and the elements $b_{j_1k_1,j_2k_2}$ have the same definition as Eq.~(\ref{eq:mu_ir}), replacing the electric dipole moment elements $\mu_k$ with the torsional $\boldsymbol{b}$-vector elements. 

Expressions to evaluate the torsional elements $\braket{v_{\tau} (k) \sigma | \hat{p}_{\gamma}e^{i n \gamma}|v_{\tau}' (k_2) \sigma}$ and $\braket{v_{\tau} (k) \sigma | e^{i n \gamma}|v_{\tau}' (k_2) \sigma}$ may be found in \citet{lankhaar:16}. 

\end{appendix}
\end{document}